\documentclass[final,3p,times,twocolumn]{elsarticle}

\usepackage{amssymb}
\usepackage[raggedrightboxes]{ragged2e}

\usepackage [english]{babel}
\usepackage [autostyle, english = american]{csquotes}
\MakeOuterQuote{"}

\usepackage{color}
\newcommand{\rev}[1]{#1}
\newcommand{\revv}[1]{#1}

\journal{Acta Astronautica}

\begin{document}

\begin{frontmatter}

\title{Cooperative Sovereignty on Mars: Lessons from the International Telecommunication Union and Universal Postal Union}

\author[inst1]{Alexander H. Ferdinand Ferguson}

\author[inst1]{Jacob Haqq-Misra\corref{cor1}}
\ead{jacob@bmsis.org}
\cortext[cor1]{Corresponding author}


\affiliation[inst1]{organization={Blue Marble Space Institute of Science},Department and Organization
            addressline={600 1st Avenue, 1st Floor}, 
            city={Seattle},
            state={WA},
            postcode={98104}, 
            country={USA}}

\begin{abstract}
As humans make ambitious efforts toward long-duration activities beyond Earth, new challenges will continue to emerge that highlight the need for governance frameworks capable of managing shared resources and technical standards in order to sustain human life in these hostile environments. Earth-based governance models of cooperative sovereignty can inform governance mechanisms for future Mars settlements, particularly regarding inter-settlement relations and the technical coordination required for multiple independent settlements to coexist. This study analyzes the International Telecommunication Union (ITU) and the Universal Postal Union (UPU), two of the oldest international organizations, which have successfully established evolving standards across sovereign nations. This analysis of the development and governance structures of these two organizations, and how they resolved key sovereignty issues, reveals principles that could be applicable to future settlements beyond Earth, particularly on Mars. Key insights include the strategic necessity of institutional neutrality, the management of asymmetric power relations, and the governance of shared resources under conditions of mutual vulnerability. The study distinguishes between a "Survival Layer" of technical standards essential for immediate safety and an "Operational Layer" governing economic and political activities, suggesting different governance approaches for each. Although some of these examples of cooperative sovereignty on Earth might not be sufficient for Mars due to its unique environment, lessons from the ITU and UPU case studies offer valuable strategies for designing flexible and sustainable governance models that can function from inception through explicit Earth-based coordination.
\end{abstract}



\end{frontmatter}



\section{Introduction}
\label{sec:introduction}

Commercial and government space agencies continue to push forward with ambitions for development of a space economy in Earth orbit, cis-lunar space, and beyond. This includes a growing population of satellites to support communication services in an expanding number of countries, the return of humans to the moon through efforts such as the Artemis mission, and exploration of possibilities for commercialization of lunar resources. Technology development is likewise underway for someday enabling cost-effective asteroid mining, which could provide a source of water for humans living in space as well as rare earth elements and precious metals for use on Earth \cite[see, e.g.,][]{elvis2021asteroids}. Mars exploration also continues to receive significant attention and investment, with national space agencies such as NASA,  the China National Space Agency, and the United Arab Emirates Mohammed bin Rashid Space Centre all intent on sending humans to Mars in the foreseeable future. Commercial space agencies are similarly contributing to this momentum\rev{; for example, SpaceX continues to advance its own technology for human spaceflight, with a long-term goal of establishing a permanent human presence on Mars \cite{musk2017making,musk2018making}. Human spaceflight is not without its risks, including the dangers of radiation to human biology, and such technological challenges could limit human futures in space. But other future projections suggest that the major technical challenges needed for} enabling humans to live in the space environment \rev{are at least theoretically} tractable \cite[e.g.,][]{zubrin2011case}. \rev{Continued} effort by commercial and government actors will reveal the possibilities and limitations of such technology for enabling long-duration human space exploration and settlement.

The ultimate success of long-duration space exploration and settlement efforts will depend on the ability to sustain human life in hostile and fragile environments, which will require novel and unprecedented advances in shared systems of governance. Problems of \rev{shared} sovereignty that arise on Earth \cite{perrez2000cooperative}, such as the need to manage common-pool resources \cite{ostrom1990governing}, will be exacerbated in space; for example, basic human needs such as breathable air will be manufactured commodities that will demand innovative management and sufficient redundancy of production in order to avoid life-threatening factors such as scarcity, price fluctuations, or monopolistic control \cite{stevens2015price}. An emerging community of scholars has already begun to discuss such problems of cooperative sovereignty and associated social, ethical, and legal challenges that are emerging or anticipated as humans continue expanding beyond Earth orbit to the moon, asteroids, and Mars \cite[e.g.,][]{milligan2014nobody,gilley2020space,froehlich2021assessing,cockell2022institutions,cockell2022interplanetary,haqq2022sovereign,nesvold2023off,schwartz2023reclaiming}. Such scholarship on the human dimensions of space exploration and settlement is a necessary complement to ongoing technological development and seeks to bridge the gap between the geopolitical realities of Earth and the range of possible solutions that could emerge through activities in space.

\rev{This scholarly debate extends to the internal political systems of potential Mars settlements. Various governance models are being proposed, ranging from direct democracy to technocracy or even authoritarian regimes, and each is justified by different perspectives on Martian conditions and historical precedents \cite[e.g.,][]{petrov2004permanent,wylie2014human,ehrman2022property,arguello2023exploration,utrata2024engineering,waller2024characterizing}. \citet{wojtowicz2021democracy} outlined three main contenders: direct democracy, technocracy, and corporate or military autocracy, and suggested that democratic or technocratic systems appear most probable for future Mars colonies. \citet{waller2024characterizing} on the other hand argued that the first permanent settlements will almost inevitably display de facto authoritarian characteristics. The analysis by \citet{waller2024characterizing} was based on the reasoning that initial Mars habitats will be small, isolated, and survival-focused, where unity of command is literally a life-or-death matter. As \citet{waller2024characterizing} noted, historical precedents and environmental constraints seem to favor an autocratic bend in the early years. \citet{wylie2014human} also emphasized that such internal governance frameworks must be systematically planned before departure, incorporating rules, norms, and stakeholder agreements into founding charters.}

\rev{While these governance models address the critical question of a settlement's internal political life, they do not solve the challenge of inter-settlement relations. Hence, this paper does not focus on the internal political life of individual settlements. Rather, we address a more fundamental challenge, from the perspective that a future Mars will not host a single colony with unified governance, but likely a collection of distinct settlements operated by different national, commercial, and scientific entities. These groups will need to coexist and cooperate on a planet with shared resources and a shared fragile environment. The central question, therefore, is not which political system will govern within a settlement, but how to create a system of cooperative sovereignty for the technical and logistical standards that enable multiple independent settlements to survive and function together.}

\rev{We proceed from the assumption that future Martian settlers, whether national or corporate, will be primarily driven by self-interest, competition, and a desire for strategic or economic advantage. We do not assume an inherent desire for equitable sharing. However, we argue that the Martian environment itself imposes a unique and brutal logic that compels cooperation on a foundational technical level. On Earth, non-cooperation on technical standards typically leads to inefficiency; on Mars, it can lead to catastrophic, mission-ending failure. The extreme hostility of the environment creates a state of enforced interdependence, where even fierce competitors are compelled to collaborate on a foundational layer of technical and safety standards to protect their own massive investments and ensure their own survival. In essence, actors are incentivized to cooperate because the cost of failure resulting from non-cooperation is orders of magnitude greater than the strategic trade-offs required to establish common protocols.}

\rev{Consider an emergency in which one settlement experiences a critical systems failure and requests immediate assistance. A neighboring settlement can only render effective aid if basic interfaces are compatible: the distress signal must be broadcast on an agreed-upon channel with a common message format, and emergency life-support systems must connect to universal ports. If any of these technical standards are misaligned, aid will be delayed or become impossible. This may lead to the loss of the entire settlement. This paper argues that organizations like the ITU and UPU provide powerful historical precedents for establishing such neutral, technical standards and protocols between independent actors without a central territorial government, and this is one of the challenges Mars settlements will face.}

\rev{The concept of cooperative sovereignty provides a framing for understanding situations in which sovereign actors must necessarily cooperate in order to achieve both self-interested and mutual goals. The term “cooperative sovereignty” was developed by \citet{perrez2000cooperative}, who stressed that “cooperation is not an independent principle supplementing state sovereignty or even a counterweight to state sovereignty, but that cooperation has to be conceived as an element of the very notion of sovereignty itself” \cite[\textit{v.,}][]{perrez2000cooperative}. This concept was applied by \citet{haqq2022sovereign} to long-duration futures on Mars by examining} models of cooperative sovereignty on Earth\rev{---such as the Antarctic Treaty System and the Convention on the Law of the Sea---that can serve as} templates, examples, analogies, or case studies \rev{(positive or negative)} for extending shared sovereignty into the space environment. Near-term ambitions in Earth orbit and cis-lunar space are already drawing upon existing precedent to develop specific policies for space; for example, the Artemis Accords are broadly based upon principles from the Outer Space Treaty of 1967, while also articulating a contemporary basis for multilateral cooperation in lunar exploration. Longer-term efforts at asteroid mining or Mars settlement face uncertain timelines, and the actual realization of such efforts could be much later than anticipated, but thinking about solutions to shared sovereignty for such long-duration missions can still provide insight on applications for shorter-term commercial and exploratory needs. In some cases, solutions for shared sovereignty on Earth can be directly translated to needs in space, but in other cases existing terrestrial models may be insufficient. It is therefore worth considering thought experiments today for positive and negative examples of shared sovereignty that could apply to space. Even if such models are never actually applied to long-duration space settlement, this exercise can yield benefits in reexamining the assumptions underlying models of shared sovereignty on Earth.

\revv{We also clarify before proceeding further that our analysis presupposes that all actors engaging in activities in space fall under the jurisdiction of existing sovereign entities on Earth. National space agencies are intended to conduct government-directed actions in space, which by definition fall under their respective national jurisdictions. Private space actors are likewise under the jurisdiction of the sovereign nation (or nations) of their incorporation, along with any other contractual obligations to sovereign entities. The Outer Space Treaty also stipulates that ``The activities of non-governmental entities in outer space, including the Moon and other celestial bodies, shall require authorization and continuing supervision by the appropriate State Party to the Treaty'' (Article VI). Our analysis is not necessarily concerned with the novel formation of sovereign entities on Mars \textit{ex nihlo}, but instead we presume that all actors seeking to operate on Mars or in space will fall under the jurisdiction of an existing sovereign entity (that is, a nation-state). Within this existing framework of sovereign entities, it remains open as to whether long-duration activities in space will be dominated by competition among sovereign entities, or cooperation. Our goal is to explore models that could foster such cooperative sovereignty on Mars.}

This study draws upon the International Telecommunication Union (ITU) and the Universal Postal Union (UPU) as examples of shared sovereignty for developing models of cooperative sovereignty in space. Both of these organizations emerged in response to historical needs for technical standardization among sovereign nations, which today function within the auspices of the United Nations and draw broad participation from a multiplicity of stakeholders. The ITU in particular provides communication standards that already apply to regions of space, which could conceivably extend to Mars either directly or by emulation. The UPU may be less directly translatable to use on Mars, but the emergence of the UPU and its present-day governance structure can provide guidance for the development and maintenance of technical standards that may arise during the process of space settlement.

\section{Case Study Overviews}

This section provides an overview of the historical development and governance structures of the ITU and UPU, which provide a background for the application of these models to space settlement. The following sections unpack the sovereignty issues underlying these models and then discuss their potential relevance to long-duration human space settlement.

\subsection{The International Telecommunication Union (ITU)}

The International Telecommunication Union was originally established in 1865 as the International Telegraph Union, when representatives of twenty European states met in Paris to attempt to organize the European telegraph network, standardize telegraphy equipment, and operate instructions across national borders \rev{(for detailed histories of the ITU, see, e.g.,} \cite{codding1994international,lyall2016international,citaristi2022international}). The ITU underwent a series of institutional and operational changes to adapt to the demands of technology. This accounted for why the organization has stood the test of time as the oldest continuous international organization in existence today.

In 1924, the International Consultative Committee for Long Distance Telephony was established to address technical and operational issues in international telephony. This committee was integrated into the ITU during the 1925 Paris Conference, marking the ITU's first foray into standards-making authority. The same conference also established the International Telegraph Consultative Committee to standardize telegraph networks.  Radio regulation became crucial due to interference issues as well as restrictions that prevented Marconi operators from communicating with non-Marconi operators. Conferences from 1903 to 1912 addressed these challenges, leading to the creation of the International Radio Consultative Committee in 1927. In 1932, the International Telegraph Convention and the International Radiotelegraph Convention were merged at the International Telegraph Conference in Madrid to form the International Telecommunication Convention. The new name, International Telecommunication Union (ITU), came into effect on January 1, 1934. The International Telephone Consultative Committee and the International Telegraph Consultative Committee were merged in 1956 to form the International Telegraph and Telephone Consultative Committee within the ITU.  On November 15, 1947, the International Telecommunication Union reached an agreement with the newly established United Nations to become a specialized agency within the UN framework. This agreement officially took effect on January 1, 1949. The integration of the ITU into the UN system occurred after World War II, during a time when there was a strong push for international collaboration to ensure peace and development. 

The need for international telecommunication standards became even more apparent in June 1965, when Intelsat made a groundbreaking achievement by launching the first international communications satellite, known as Early Bird or Intelsat I. This satellite was positioned in a geostationary orbit over the Atlantic Ocean, initially connecting North America and Europe, and subsequently extending its reach to other countries on both sides of the Atlantic. Managed by Comsat, Early Bird was the pioneer in providing commercial telephone and television services via satellite. Two years later, in 1967, Intelsat expanded its network by placing Intelsat II over the Pacific Ocean, thereby linking the Pacific regions with the Atlantic. By 1969, the addition of Intelsat III over the Indian Ocean completed the network, achieving comprehensive global coverage. 

In 2006, the World Telecommunication and Information Society was established to commemorate the establishment of the ITU. Today, ITU currently coordinates the global use of the radio spectrum, assignment of geostationary satellite orbits, development technical standards for telecommunications, and efforts to improve access to Information and Communication Technologies (ICTs) in underserved communities worldwide---particularly in developing countries---in order to bridge the digital divide through policy and regulatory activities and setting global best practices for ICT services.

The ITU's role in facilitating global communication has made it a critical component of the international framework for cooperation. Comprising 193 member states and over 900 industry members today, the ITU focuses on setting global standards and managing radio frequencies, which are crucial for maintaining efficient and interoperable communication systems worldwide. The ITU consists of the following: (1) the Plenipotentiary Conference; (2) World Administrative Conferences; (3) the ITU Council; (4) the General Secretary; (5) Radiocommunications Sectors; (6) Telecommunications Standardization Sectors; and (7) Telecommunications Development Sector. The organization has evolved since its inception to address new technologies, including telephony, wireless telegraphy, and satellite communications, which has played a crucial role in promoting international cooperation and standardization in the field of telecommunications.

The ITU uses a hybrid model of governance which combines intergovernmental decision making with private sector participation. The organization has a policy-making body, the Plenipotentiary conference, which meets every four years to determine ITU policies and strategies, adopt strategic and financial plans and elect its management. The ITU council acts as ITU’s governing body. It has 48 member states elected by Plenipotentiary conference and meets annually. The Secretary-General is elected by the Plenipotentiary conference for 4-year terms and manages the organization as its chief executive. The ITU has three main sectors: Radiocommunication (ITU-R) which manages radio frequency spectrum and satellite orbits; Standardization (ITU-T) which develops technical standards; Development (ITU-D) which provides technical assistance to developing countries. For decision making, the ITU mostly uses a consensus-based approach such as one country, one vote system for member states. Private sector members can participate but not vote. Recommendations of the ITU may become binding if adopted by member states.

The ITU's regulations are primarily designed to manage spectrum use originating from Earth or affecting Earth-based services. Communications solely occurring on or around Mars, without impacting Earth, fall outside the ITU's current regulatory scope. However, the ITU is currently involved in regulating and coordinating radio frequencies for space services, including those used in lunar missions. At the recent World Radiocommunication Conference (WRC-23), lunar communications were added to the agenda for the next conference in 2027. This is intended to establish a framework for managing the radio spectrum for communications on the lunar surface and between lunar orbit and the surface.

\subsection{The Universal Postal Union (UPU)}

The Universal Postal Union (UPU) sought to establish uniform postal rates and ensure equal treatment for domestic and international mail. Heinrich von Stephan, together with others, played a crucial role in the founding of the UPU. As the General Post Director for the German Empire, he organized the International Postal Congress in Bern in 1874, where he successfully gathered postal representatives from 22 countries. His proposals led to the signing of the Treaty of Bern on October 9, 1874, which established the General Postal Union, later renamed the Universal Postal Union in 1878. Von Stephan's efforts were instrumental in creating a single postal policy that simplified and standardized international mail services, introducing uniform postal rates, and ensuring equal treatment for domestic and international mail. The UPU operated independently until 1948, when it became a specialized agency of the United Nations. Initially, The UPU's host country, Switzerland, actively resisted pressure to join the League of Nations (the UN's predecessor) after World War I, aiming to protect the UPU from political influence. But the UPU’s incorporation into the UN was generally accepted as positive and seen as a way to strengthen international postal cooperation, especially after World War II; thus most UPU members viewed it as a beneficial move to align with the broader international framework provided by the UN. This integration allowed the UPU to contribute to UN policies related to social and economic development. In recent years, the UPU has overseen initiatives such as the Express Mail Service cooperatives, which facilitate international express mail delivery. \rev{(For a detailed history of the UPU, see, e.g.,} \cite{lyall2016international}).

The governance structure of the UPU consists of four main bodies: the Congress, the Council of Administration, the Postal Operations Council, and the International Bureau. The Congress is the supreme authority of the UPU and meets every four years to examine proposals to amend UPU acts, including the Constitution, General Regulations, Convention, and Postal Payment Services Agreement. It serves as a forum for discussing issues impacting international postal services. The Congress itself is composed of representatives from member countries, and these representatives are usually appointed by their respective governments. The elected positions within the UPU, such as the Director General, usually have a term length that aligns with this four-year cycle. The UPU's decision-making process, just like the ITU, is based on the principle of one country, one vote. This approach ensures equal representation for all member states but may not always align with the economic realities of the global postal market. The organization has grappled with this issue, particularly in debates over terminal dues and market-dominant positions of certain postal operators. 

The Council of Administration and the Postal Operations Council carry out the day-to-day governance of the UPU. This system separates administrative and operational functions, allowing for more focused and efficient management. However, this separation also raises questions about coordination and the potential for conflicting priorities. The UPU does not have direct regulatory authority over private postal carriers. These companies operate under commercial agreements and are subject to the regulations of the countries in which they operate. However, the UPU does engage with private sector stakeholders through its Consultative Committee. This committee includes industry associations representing private postal operators, express carriers, manufacturers, and e-commerce companies. The UPU structure allows for technical expertise to inform decision-making, while still preserving the intergovernmental nature of the organization. However, the organization continues to face challenges in adapting to the evolving landscape of global communications and commerce, particularly in the digital age.

\section{Cooperative Sovereignty under the ITU and UPU}

\rev{This section examines how the ITU and UPU navigated critical sovereignty challenges throughout their development. Our analysis focuses on the strategic necessity of neutrality, the management of asymmetric power relations, and the governance of shared resources under conditions of mutual vulnerability which were common recurring patterns. Each of these patterns offers insights for future Mars settlements facing similar tensions between competition and cooperation.}

\subsection{Strategic Neutrality for Network Value}\label{sec:swiss}

\rev{The first conference of the ITU was held in Paris in 1865, and there was an initial suggestion that the Union's headquarters be located in the city where the most recent conference had taken place. However, this proposal was swiftly rejected due to widespread concern that locating the headquarters in Paris would place the organization under the influence of France, then led by Napoleon III. While the stated aim was to ensure the Union's independence and uphold the principle of `equal sovereignty among nations,' the underlying incentive was more of preserving the value of the network itself (discussed further by \citet{codding1994international}).}

\rev{This is because, for the new telegraph union to be economically viable, it needed participation from all major European powers. The ITU being dominated by France would likely have been perceived by rival powers like the Great Britain as a geopolitical tool of Paris. This perception could have led them to either refuse to participate or establish a competing, incompatible network, which will result in division and a significantly less valuable system for all the parties involved. The major member states therefore agreed on a politically neutral location in Switzerland as it was the only rational choice to maximize their individual and collective interests. Neutrality was the necessary concession that was used to create a unified, functional, and globally valuable communications network (see, e.g., the analysis by \citet{balbi2013swiss}).}

\rev{This precedent reveals that when the value of a system depends on universal participation, even the powerful and dominant actors will accept institutional constraints to prevent fragmentation. This suggests that on Mars, settlements will need to establish governance mechanisms that are institutionally separated from any single actor's control, even if that actor provided the initial infrastructure. For instance, this could be in the form of rotating leadership or establishing physically neutral facilities to house coordinating functions.}

\subsection{Asymmetric Power Relations}\label{sec:power}

\rev{The ITU's evolution also shows challenges arising from asymmetric power relations. The organization transitioned from being dominated by developed countries to including a growing number of developing nations. The developing countries requested the creation of a special fund to provide technical and developmental assistance, to be financed from the ITU budget. However, these requests were not fulfilled in their original form. Instead, a development assistance fund was established through voluntary contributions rather than mandatory budget allocations. This compromise, in a way, preserved the developed nations' financial control while providing some support to the newcomers.}

\rev{Early settlers on Mars will almost certainly come from technologically advanced, wealthy nations or well-capitalized corporations. The immense technological and financial barriers to entry mean any initial group on Mars will be highly sophisticated. However, this does not eliminate the potential for significant power imbalances to emerge. The ITU's "developed vs. developing" countries dynamic could play out in the form of "Established Pioneers vs. New Arrivals" on Mars.}

\rev{The first entity that will establish a large-scale, self-sustaining presence will hold immense advantages. They will control the most valuable resources, possess established infrastructure (landing pads, power grids, communication relays), and dictate the initial technical standards through their operational choices. A new settlement that arrives years later, even if well-funded, will face significant disadvantages, becoming a "developing" entity within the Martian context.}

\rev{The ITU experience gives a clue why established powers cannot simply dictate terms indefinitely. A dominant settlement cannot function in isolation; it needs other settlements to recognize its standards and participate in its systems. Mars creates some shared vulnerabilities that could align the interests of established and new actors around certain baseline standards. Unlike Earth, the greatest existential threat on Mars is a shared one; any new settlement that cuts corners on planetary protection protocols, life-support safety, or waste management could create a catastrophe that affects everyone. Newcomers can offer their verifiable commitment to uphold critical, shared-risk standards thereby protecting the investment and safety of established players.}

\rev{Additionally, established powers may possess strengths in some areas while facing critical gaps in others. For instance, an established mining consortium might be rich in engineers but poor in biological expertise. A newer scientific outpost, while resource-constrained, could bring invaluable human capital needed for long-term food security, medical research, or environmental monitoring, capabilities the established power may need to contract for. This could create mutual dependencies that incentivize cooperation despite power asymmetries.}

\subsection{Spectrum Management}

\rev{The ITU's management of the radio spectrum provides a useful analogy for Mars governance. The radio spectrum is an example of a finite, shared resource that, without regulation, becomes useless to everyone because individuals acting in their own self-interest could interfere with one another. On Earth, this kind of interference is primarily an economic and logistical problem, causing service disruptions and commercial losses. However, on Mars it becomes a life-or-death crisis.}

\rev{A single instance of signal interference at a critical moment could lead to a crashed lander (destroying a billion-dollar asset and killing the crew), a lost astronaut who cannot transmit distress coordinates, or the failure of automated life-support systems to receive critical commands. The ITU's Radio Regulations represent a binding treaty of mutual restraint that is born from exactly this kind of necessity. Member states agree to cede a portion of their sovereignty, in this case the absolute freedom to broadcast on any frequency, in exchange for the far greater benefit of predictable, reliable, and functional access to the spectrum for all.}

\rev{For Mars, this logic becomes even more compelling. Competing settlements will be compelled to establish an ITU-like regulatory body for the Martian spectrum because the alternative is a communications environment that is so chaotic that it jeopardizes the survival of every human on the planet. The governance of the spectrum becomes a foundational layer of order that must be established before any other activity, competitive or cooperative, can even be considered. This is what we consider the "Survival Layer" of Mars governance: the set of technical standards that are so critical to basic safety that even the most antagonistic competitors will cooperate to maintain them.}

\rev{It is worth noting that the spectrum allocation challenge is also time sensitive. Unlike other disputes such as resource extraction that can be negotiated gradually, spectrum interference has immediate consequences. This suggests that Mars settlements will need to establish basic communication protocols and frequency allocation systems before significant multi-settlement activity occurs, rather than waiting for conflicts to emerge.}

\subsection{The UPU and the Issue of Economic Integration}

\rev{Moving from the ITU to the UPU, we encounter similar sovereignty challenges resolved through different mechanisms. The case of Japan's entry into the UPU in the late 19th century shows that technical standardization can serve as a pathway to political recognition and economic integration, even for previously marginalized actors.}

\rev{In the mid-19th century, Western nations imposed "unequal treaties" on Japan to expand their economic and strategic interests in East Asia. These treaties, initiated after Commodore Matthew Perry's "Black Ships," forced Japan to open to the world in 1853 and were characterized by significant disparities in power between Western powers and Japan. This led to Japan's loss of sovereignty and autonomy. These arrangements were not necessarily responses to Japanese aggression, but possibly the Western desires to open new markets, and extend geopolitical influence.}

\rev{Japan sought to repeal these unequal treaties by demonstrating it could function as an equal partner in international systems, and joining the Universal Postal Union became a key strategy in this campaign. One of the major incentives for the Western powers to eventually accept Japan as an equal partner in the UPU was economic efficiency. An unequal, divisive postal system with multiple sets of rules will be complex, slow, and detrimental to international commerce. The Western nations gained reliable and cheap access to the Japanese market once Japan was integrated into the standardized UPU system. It becomes clear that when exclusion from standardized systems becomes more costly than inclusion, even dominant powers will grant recognition to weaker actors.}

\subsection{The 2018 UPU Terminal Dues Crisis}\label{sec:terminaldues}

\rev{The UPU's terminal dues system, established in 1969, provided deep discounts for businesses in China and other developing countries shipping products into the United States. By the 2010s, this system was seen as distorting global package flows, incentivizing the shipping of goods from foreign countries that benefited from these discounts and creating unfair competitive advantages for foreign mailers over domestic ones. In 2018, the United States threatened to withdraw from the UPU due to concerns that the existing terminal dues structure unfairly advantaged certain developing nations, particularly China.}

\rev{For the US, the rational choice was not simply to exit. A full withdrawal would have created immense logistical chaos for American businesses dependent on international mail services. Its strategy was to use the credible threat of withdrawal to force a renegotiation that better served its interests. The other UPU members also faced a dilemma of whether they could refuse US demands on principle, to preserve the existing "equitable" system, even though this move could risk the collapse of the entire global postal network upon the exit of its largest member.}

\rev{The UPU case also reveals a critical limitation; the willingness to compromise depends on all parties recognizing that the alternative (system fragmentation) is worse than making concessions. On Mars, this calculation holds only as long as settlements genuinely need each other. If technological advancement ever enables true self-sufficiency, the incentive structure for cooperation could collapse.}

\section{Lessons for Mars Settlement}

The existing ITU and UPU structures can provide both direct and indirect models for application to long-term Mars settlement as well as shorter-term ambitions for lunar or asteroid commercialization. The ITU itself is already assessing its ability to enact lunar telecommunication standards, and future developments could conceivably extend the ITU’s jurisdiction to activities on asteroids or even on Mars. But it also remains possible that the domain of the ITU will remain restricted to transmissions that originate on Earth: these could potentially apply to Earth-to-Mars transmissions, but telecommunication systems that are developed by future Martian settlers, for use within Mars only, may be beyond ITU control. At the same time, the development of any technical standards on Mars, whether for telecommunication or other purposes, will raise novel challenges that would not arise on Earth. 

\rev{We started our analysis with the following assumptions: a future Mars will not be a monolith controlled by a single government or corporation. Instead, it will host a collection of distinct settlements operated by different national, commercial, and scientific entities, each with its own self-interest. This landscape is the necessary precondition for inter-settlement relations and creates the interoperability challenges the ITU and UPU were built to solve. As we pointed out earlier, no single entity currently possesses the political, economic, and technological monopoly required to claim, colonize, and exclude all others from an entire planet.}

\rev{While a single dominant actor is possible for a time, our argument still holds in two ways. As the entity's presence grows, it will form specialized divisions which will require rigid, standardized technical protocols to interact, manage resources, and ensure mutual safety. So, there will still be the need for cooperation at an internal corporate or state-level to prevent one division's failure from wiping out the entire enterprise.}

\rev{In addition, Mars settlements, even a "totalitarian" one, will not exist completely in isolation especially because future Martian settlers would originate from Earth and therefore will likely interact with Earth, lunar settlements, or asteroid miners to trade, receive supplies, or manage communications. To do so, it will have to adopt common technical standards for communication and transport, just as 19th-century Japan, despite being isolationist for some time, was rationally incentivized to join the UPU to gain access to the global economic network.}

\subsection{Technical Standards as Foundational Necessity}

Consider \revv{a} scenario where two settlements A and B, \revv{both under the jurisdiction of existing sovereign nations,} exist on Mars simultaneously. \revv{(One example of such a scenario could be a near-simultaneous attempt by both NASA and the China National Space Administration to establish a human presence on Mars.)} Before they decide whether to cooperate or compete, they face an immediate technical problem: their radio transmissions can interfere with each other. If Settlement A is attempting a resupply landing while Settlement B is conducting an extravehicular activity, and both are using overlapping frequencies, the result will be catastrophic. This is what we mean by the Survival Layer: the set of technical standards that must function correctly for anyone to survive long enough to engage in \revv{competition or cooperation} at all. The Survival Layer is the precondition for \revv{mutual interaction between sovereign actors}.

\rev{The ITU demonstrates this logic in part. When the European powers established the International Telegraph Union in 1865, they did not \revv{wait to} resolve their territorial disputes or ideological differences \revv{before cooperating on technical matters}. They coordinated telegraph standards because without standardization, the telegraph network had no value to anyone. On Mars, this becomes absolute. It is impossible to negotiate resource extraction treaties if communication systems do not work. It is impossible to establish trade agreements between habitats that use incompatible life-support interfaces. It is even impossible to engage in competition if the basic infrastructure of survival is fragmented and dysfunctional.}

\subsection{The Operational Layer and Its Different Logic}

\rev{While the Survival Layer demands rigid, universal standards that must be enforced through mutual vulnerability, governance challenges beyond immediate survival require different approaches. We term this the Operational Layer: the realm of economic activity, resource extraction, scientific research, and internal political organization within settlements.}

\rev{The Operational Layer decisions can afford experimentation, competition, and diversity because failures in this domain, while costly, are not immediately catastrophic to all settlements. If one settlement adopts an inefficient economic model or an unpopular internal political system, it harms primarily itself. This creates space for the kind of governance diversity discussed in the literature on internal Mars political systems \cite{wylie2014human,utrata2024engineering,wojtowicz2021democracy}.}

\rev{However, the Survival and Operational Layers are not entirely separate. Operational decisions can cascade into Survival Layer crises if they compromise critical safety margins. For example, economic desperation might lead a settlement to cut corners on life-support redundancy, increasing the probability of catastrophic failures requiring emergency assistance. This follows that there should be a two-tier governance model for Mars, one that is rigidly standardized and internationally coordinated protocols for the Survival Layer, combined with a flexible, adaptive arrangement for the Operational Layer. The ITU and UPU provide templates for the former but offer limited guidance for the latter.}

\subsection{Why Cooperation Will Persist Despite Competition}

\rev{A critical question is whether cooperative governance on Mars can survive long-term competitive pressures. Earth history is filled with examples of international cooperation collapsing under geopolitical rivalry. What makes Mars different? The UPU terminal dues crisis (\S\ref{sec:terminaldues}) offers a good lesson that cooperative institutions survive only when all parties calculate that fragmentation is worse than compromise. On Earth, the United States could credibly threaten the UPU with withdrawal because alternative arrangements were feasible, if inconvenient. The threat worked because exit, even though costly was very possible.}

\rev{On Mars, this changes fundamentally. On Earth, postal system fragmentation means slower mail and higher costs. On Mars, communication or life-support standard fragmentation means astronauts could die and settlements could get destroyed. The difference between Earth inconvenience and Mars’ capacity to cause significant instant damage makes defection from Survival Layer standards very costly, even for powerful actors.}

\rev{However, this guarantees cooperation only during a limited historical window. If technological advancement eventually enables true long-duration settlement with full self-sufficiency (e.g., closed-loop life support, complete resource independence, sufficient population for specialized expertise), then the incentive structure could collapse.}

\subsection{Institutional Design Principles for Mars Governance}

\rev{Drawing on the ITU and UPU experiences, we propose five design principles for Mars governance institutions:}

\begin{enumerate}
    \item \textit{Institutional Neutrality by Design} -- \rev{The ITU's Swiss headquarters precedent (\S\ref{sec:swiss}) demonstrates the necessity of perceived neutrality. On Mars, any inter-settlement coordinating body must be institutionally separated from individual settlement control. These could include rotating leadership among settlement representatives and having multi-stakeholder boards with weighted voting that prevents any single actor from dominating.}
    \item \textit{Flexibility and Renegotiation Mechanisms} -- \rev{The UPU terminal dues crisis (\S\ref{sec:terminaldues}) shows that rigid institutional structures cannot survive changing power dynamics. Mars governance must include explicit mechanisms for periodic review and adjustment of standards (e.g. conferences similar to ITU World Radiocommunication Conferences).}
    \item \textit{Graduated Participation Requirements} -- \rev{The ITU's voluntary development fund (\S\ref{sec:power}) suggests a model for managing power asymmetries. Mars governance could implement mandatory compliance with Survival Layer standards (non-negotiable for any settlement) and voluntary participation in Operational Layer coordination (economic arrangements, research sharing).}
    \item \textit{Enforcement Through Mutual Vulnerability Rather Than Sanctions} -- \rev{Traditional international organizations rely on sanctions, diplomatic pressure, or military enforcement. Mars settlements lack these mechanisms. Enforcement could come automatically through mutual vulnerability; a settlement that broadcasts on unauthorized frequencies will be immediately cut off by others in self-defense. A settlement with incompatible life-support interfaces will not receive emergency aid.}
    \item \textit{Earth-Based Precedent-Setting} -- Unlike the ITU and UPU, which emerged gradually through trial and error, Mars settlements require functional governance from inception. This demands unprecedented Earth-based coordination among spacefaring \revv{sovereign} nations and corporations \revv{under their jurisdiction} to establish Survival Layer standards before departure. Some possible mechanisms could include expanding the Artemis Accords framework to include Mars-specific technical standards and requiring corporations seeking Mars mission licenses to demonstrate compliance with internationally recognized safety protocols.
\end{enumerate}

It is important to note that the ITU governs thousands of entities across billions of people while early Mars settlements will involve handfuls of outposts and hundreds of individuals. Governance structures must scale appropriately without imposing the Earth-level bureaucratic overhead. Also, terrestrial conflicts will inevitably extend to Mars. If Earth-based rivals (e.g., US--China tensions) refuse cooperation, even compelling survival logic may prove insufficient. The ITU/UPU functioned during Cold War tensions, but Mars's smaller scale and higher stakes could make rivalries more volatile.

\section{Conclusion}

\rev{The International Telecommunication Union and Universal Postal Union demonstrate that cooperative sovereignty among competitive actors is achievable when the value of universal participation outweighs the benefits of unilateral action. Their histories reveal some interesting insights for future Mars governance.}

\rev{First, technical standards become primary governance challenges when survival depends on their universal adoption. The ITU's spectrum management and the UPU's postal integration show that even geopolitical rivals will cooperate on foundational systems that cease to function without universal participation. On Mars, this logic will intensify: communication failures, incompatible life-support systems, and inadequate safety protocols become existential threats rather than mere inconveniences.}

\rev{Second, cooperation persists despite power imbalances when shared vulnerabilities align interests. The ITU's evolution from developed-country dominance to inclusive global governance shows that established powers will accommodate newcomers when exclusion threatens system stability. The UPU's terminal dues renegotiation shows that rigid frameworks collapse under changing conditions. Mars governance must anticipate power shifts, technological changes, and evolving needs through explicit review mechanisms and amendment procedures.}

\rev{However, the ITU and UPU also reveal a limitation that cooperative institutions endure only during windows of enforced interdependence. If Mars settlements eventually achieve true self-sufficiency, the compulsion for cooperation could erode. }

\section*{Acknowledgments}
\revv{Thanks to two anonymous reviewers for providing feedback that significantly improved this paper.} This study was conducted during the 2024 Young Scientist Program (YSP) at the Blue Marble Space Institute of Science. This research did not receive any specific grant from funding agencies in the public, commercial, or not-for-profit sectors. Any opinions, findings, and conclusions or recommendations expressed in this material are those of the authors and do not necessarily reflect the views of any employer.

\bibliographystyle{elsarticle-num-names} 
\bibliography{refs}

\end{document}